\documentclass[10pt,aps,pra,twocolumn,showpacs,superscriptaddress]{revtex4-2}
\usepackage[english]{babel} 	
\usepackage[usenames, dvipsnames]{color} 
\usepackage{graphicx} 
\usepackage{bm} 
\usepackage{amsmath} 
\usepackage{amsthm} 
\usepackage{amssymb} 
\usepackage{times} 
\usepackage[pdftex,colorlinks=true,urlcolor= blue,linkcolor=Blue,citecolor=RedViolet]{hyperref}
\usepackage{multirow}
\usepackage{physics}

\providecommand{\bra}[1]{\langle #1 |}
\providecommand{\ket}[1]{| #1 \rangle}

\usepackage[small,bf]{caption}

\usepackage{soul}

\begin{document}

\title{System-environment quantum information flow}

\author{Taysa M. Mendon\c{c}a}
\email{tmendonca@ifsc.usp.br}

\affiliation{Instituto de F\'{i}sica de S\~{a}o Carlos,
	Universidade de S\~{a}o Paulo, CP 369, 13560-970, S\~{a}o Carlos, SP, Brazil}
\affiliation{Centre for Quantum Materials and Technologies, School of Mathematics and Physics, Queen’s University, Belfast BT7 1NN, United Kingdom}

\author{Lucas C. C\'eleri}
\email{lcceleri@ufg.br}
\affiliation{QPequi Group, Institute of Physics, Federal University of Goi\'as, Goi\^ania, GO, 74.690-900, Brazil}

\author{Mauro Paternostro}
\email{mauro.paternostro@unipa.it; m.paternostro@qub.ac.uk}
\affiliation{Universit\`a degli Studi di Palermo, Dipartimento di Fisica e Chimica - Emilio Segr\`e, via Archirafi 36, I-90123 Palermo, Italy}
\affiliation{Centre for Quantum Materials and Technologies, School of Mathematics and Physics, Queen’s University, Belfast BT7 1NN, United Kingdom}

\author{Diogo O. Soares-Pinto}
\email{dosp@ifsc.usp.br}

\affiliation{Instituto de F\'{i}sica de S\~{a}o Carlos,
	Universidade de S\~{a}o Paulo, CP 369, 13560-970, S\~{a}o Carlos, SP, Brazil}

\begin{abstract}
	
	
	
	The return of the information from the environment to the system is a phenomenon can be related to existence of non-Markovian mechanisms in the environment and such transformation of resources can be useful for quantum information applications. Thus, understanding the details of the system-environment information dynamics, i.e., the transference of quantum resources is of key importance to design noise-resilient quantum technologies. In this work, we show how a quantum resource propagates from the main system to an environment, using as model a single qubit coupled to two linear chains of qubits, and also the information dynamics among the environment qubits. In this way, we characterize the propagation of information leaving the main qubit and going through the environment. Finally, we connect the conditions for the emergence of this dynamics to the existence of quantum Darwinism.
	
\end{abstract}

\pacs{Pacs}

\maketitle

\section{Introduction}

{In an open quantum system, a backflow of information from the environment can take place, thus giving origin to non-Markovian dynamics. Such feature can be exploited in a number of quantum tasks, including in the emerging domain of quantum thermodynamics~\cite{Ghosh2018,Abiuso2019,Serra2020a,Paternostro2020}. 
	Characterizing the information flow entailed by non-Markovian dynamics is thus important both fundamentally and practically.}

Various quantities have been proposed to quantitatively address the non-Markovian behavior of a given quantum evolution \cite{Rivas2014,Colloquium_Vacchini}. Among them, the one proposed by Breuer, Laine and Piilo (BLP) ~\cite{Breuer2009,Breuer2010} associates non-Markovianity to the lack of dynamical state-homogeneization: the distinguishability of two initial states of a quantum system -- as measured  by the trace norm -- is a contractive function of time under a Markovian quantum channel, signalling asymptotic state-homogeneization. 
Thus, a phenomenology resulting in the violation of such monotonicity can be interpreted as a signature of non-Markovianity, and used to quantify the {\it degree}  of departure of a given evolution from the Markovian framework. 
A significant body of literature (both theoretical and experimental) on this subject has been produced to date, and much has been established of  the mathematical foundations and applications of non-Markovianity \cite{Rivas2014,Colloquium_Vacchini, Dariusz2022, Li2018, Vega2017, Whitney2018, Ghosh2018, Sampaio2017, Bhattacharya2017, Gelbwaser2013, Maniscalco2018, Serra2020a}. However, few were the studies about the information flow between system and environment and details of the mechanisms of creation or destruction of quantum resources \cite{santos2024quantifying, Vega2017, Paternostro2014c, Paternostro2019, Maziero2010}. 

Establishing the formalism able to describe such flow and identifying instances of physical situations that allow for its quantitative assessment would be very important. Besides shedding further light on the phenomenology of non-Markovian quantum dynamics, it  
will enable the identification of the fundamental mechanisms leading to the quantum-to-classical transition, as established -- for instance -- by the framework of Quantum Darwinism~\cite{Zurek2009_Darwin, Zurek2013_Darwin}. There, the quantum information shared by different observers is central to the  assessment of the process leading the quantum state of a system to classicality through the emergence of redundant information encoding: when the same information is spread across the elements of a multi-partite environment, and can be collected by associated observers, classicality is present~\cite{Zurek_DarwinDeco,Zurek2013_Darwin, Zurek2009_Darwin}. The emergence of Quantum Darwinism can thus be signalled by quantifying mutual information between the main system and growing-in-size subparts of the environment \cite{Zurek_DarwinDeco}.
Several dynamical models have been studied theoretically through the Quantum Darwinism framework~\cite{Castro2019_StrongDarwin, Castro2020_StrongDarwin2,Mauro_Darwin2021,Mauro2022_Darwin2, Darwin_Sebastian2023, Mauro2019_Darwin1} and on experimental platforms \cite{Mauro2018_DarwinExp, Chen2019_DarwinExp, Jelezko2019_DarwinExp}. The key nature of Quantum Darwinism as a diagnostic tool for the understanding of how quantum information flows from the system to the environment makes it well suited to study quantum non-Markovianity.


In this work we use the BLP measure~\cite{Breuer2009,Breuer2010} to study the  link between non-Markovianity of the dynamics of a quantum system and the flow of information to a finite-size environment. Remarkably, we are able to characterize such flow fully through the degree of initial quantum coherence in the state of the system.
We 
study the conditions for information trapping within the environment, and return to the system.  
The assessment of dynamically-created coherence, and the propagation of information from the system, paves the way to the study of the occurrence of Quantum Darwinism.
We show that the instants of time when information is sent to the environment or returns from it, correspond to when equal amounts of information are present across different fragments of the environment, thus identifying unquestionably the manifestation of quantum Darwinism. 

\noindent
{\it Model and resulting dynamics.--}
{We take inspiration from the quantum  model proposed in Ref.~\cite{Mendonca2020}, consisting of a single qubit system coupled to a finite-size environment, to construct a mechanism for open system dynamics where information travels through the environment in discrete time steps.
	We thus consider the adamantane molecule (C$_{10}$H$_{16}$) in the presence of a strong static magnetic field as the {\it platform} for our investigation. The compound consists of six $\text{CH}_2$ and four CH groups. The carbon spin from the $\text{CH}_2$ group plays the role of a system qubit, which is coupled to  an environment comprising two linear chains of $N$ qubits each. Such configuration has been realized experimentally in a nuclear magnetic resonance (NMR) platform~\cite{Mendonca2020,Alvarez2010,Ajoy2011,Souza2011}.
	The total number of qubits in the full system is then $2N + 1$, as sketched in Fig.~\ref{fig:Sistema}.}


{Ref.~\cite{Mendonca2020} showed that the nuclear spins of the hydrogen in the molecule can be considered as a thermal bath for the nuclear spin of a single $^{13}$C atom, which embodies the main system (we have approximately one $^{13}$C nuclear spin for 160 $^1$H spins, which thus represent a spin bath for the carbon nuclear spin). {The natural Hamiltonian of the compound reads~\cite{SlichterLivro1990} $H^{(0)} = H^{S}_{z}+H^{E}_{z}+ H^{(0)}_{SE} + H^{(0)}_{E},$ 
		where $H^{S}_{z}$ and $H^{E}_{z}$ represent the Zeeman energies induced by the interaction of system and environment with the external magnetic field, respectively. They can be discarded in a suitable rotating frame at the Larmor frequencies associated with such Zeeman terms ~\cite{Alvarez2010,Ajoy2011,Souza2011}. 
		The corresponding Hamiltonian reduces to a term describing the system-environment coupling, and one accounting for the interaction among the elements of the environment, and can be written as} 
	$H = H_{SE} + H_{E}$ with 
	(we assume units such that $\hbar = 1$ throughout the text)~\cite{Mendonca2020}
	\begin{equation}
		{\displaystyle H_{SE}=}	J_{SE}\sum_{\alpha=a,b}\left(2\sigma_{z}\varepsilon_{z}^{\alpha,1}+\sigma_{x}\varepsilon_{x}^{\alpha,1}+\sigma_{y}\varepsilon_{y}^{\alpha,1}\right),
	\end{equation}
	and
	\begin{equation}
		H_{E}={J_{E}}\sum_{\alpha=a,b}\sum_{k=1}^{N-1}\left[2\varepsilon_{z}^{\alpha,k}\varepsilon_{z}^{\alpha,k+1}{-}\left(\varepsilon_{x}^{\alpha,k}\varepsilon_{x}^{\alpha,k+1}+\varepsilon_{y}^{\alpha,k}\varepsilon_{y}^{\alpha,k+1}\right)\right].
	\end{equation}
	Here, $\sigma_{\mu}$ and $\varepsilon_{\mu}^{\alpha,k}$ $(\mu=x,y,z)$ are the system and environment spin operators, respectively. They satisfy (anti-)commutation relations of spin-1/2 particles and are thus akin to Pauli matrices. The label $\alpha$ identifies the environmental chains, whose elements are labelled  as $k=1,\dots,N-1$. The coupling rate between the main system and the first qubit of each chain is denoted by $J_{SE}$, while $J_{E}$ stands for the interaction rate between neighbouring qubits in each chain.
	
	\begin{figure}[t]
		\includegraphics[width=0.7\columnwidth]{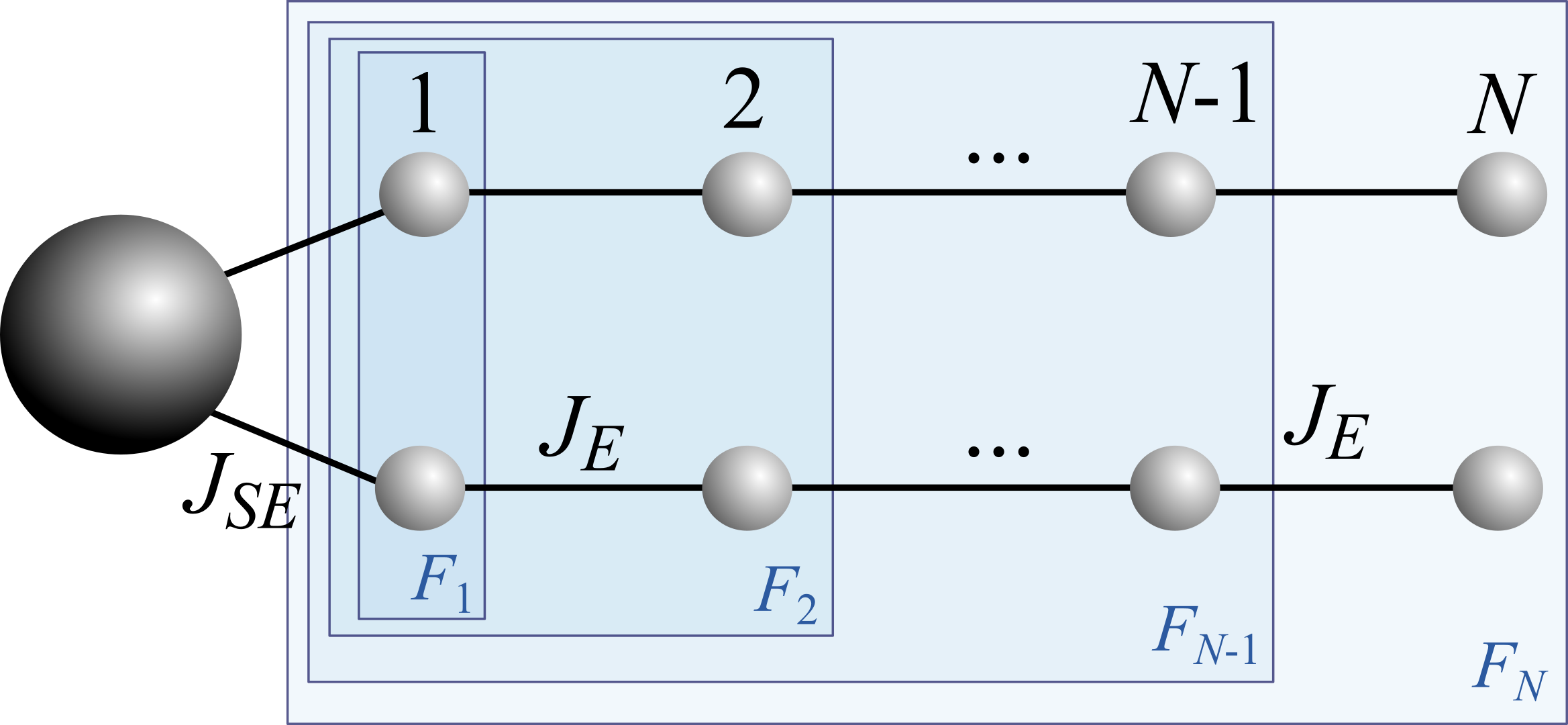}
		\caption{{Sketch of the physical situation being considered}. The model consists of a system qubit (dark grey dot) coupled to an environment comprising two distinct chains of qubits (small bright dots). Each blue-colored rectangle refers to the $m^\text{th}$ fragment $F_m$ of the environment. Here, $J_{SE}$ ($J_E$) is the coupling rate between the system and the first element of each chain (the intra-chain coupling rate).}
		\label{fig:Sistema}
	\end{figure}
	
	In this context, we are interested in characterizing the dynamics of the main system. It is know that a system coupled to an environment with finite degrees of freedom can present a non-Markovian behaviour \cite{Breuer2016_colloquium} and one particular form to quantify such characteristic is based on the indistinguishability of quantum states, as measured by the trace distance~\cite{Breuer2009,Breuer2010}. As such quantity is contractive 
	under Markovian evolution, its non-monotonic behaviour signals non-Markovianity stemming from a flow of information from the environment back to the system. In this light, we will follow the dynamics of the trace distance between two distinct states of the system and identify the cases in which this quantity is non-monotonic over time.
	
	Let $S$ be prepared in the initial states 
	$\rho_{S}^{(\pm)} (0)=\ket{\pm}\bra{\pm}_S$ with $\ket{\pm}_S=(\ket{0}\pm\ket{1})_S/\sqrt{2}$ and $\{\ket{0},\ket{1}\}$ embodying the computational basis of our problem {(we have shifted the energy of such states in a way that $\sigma_z\ket{0}=\ket{0}$ and $\sigma_z\ket{1}=-\ket{1}$)}.
	The initial state of the environment is $\rho_{E}(0)=\bigotimes_{\alpha=a,b}\bigotimes^{N}_{k=1}\ket{0}\bra{0}_{\alpha,k}$ and we consider uncorrelated system-environment initial states in the form $\rho_{SE}^{(\pm)}(0)=\rho_{S}^{(\pm)}(0)\otimes\rho_{E}(0)$. The  evolution is governed by the Liouville-von Neumann equation 
	\begin{equation}
		\label{LvN-eq}
		\frac{d}{dt}\rho_{SE}^{(\pm)}(t)= - i\left[H, \rho_{SE}^{(\pm)}(t)\right],
	\end{equation}
	which can be solved to find $\rho_{SE}^{(\pm)}(t)$ and, in turn,  the reduced states of the system $\rho_{S}^{(\pm)}(t)$. The distinguishability between such states is then defined as 
	\begin{equation}
		\label{TD_sisAmb}
		D_{S}\left(\rho_{S}^{(+)}(t),\rho_{S}^{(-)}(t)\right)= \left\|\rho_{S}^{(+)}(t)-\rho_{S}^{(-)}(t)\right\|_{1}
	\end{equation}
	with $ \| \text{A}\|_{1} = \frac{1}{2}\mbox{Tr}\left(\sqrt{\text{A}^{\dagger}\text{A}}\right)$. Such quantity decreases monotonically under Markovian dynamics, which implies 
	\begin{equation}\label{sigma}
		\sigma_{S}(t)=\frac{d}{dt}D_{S}\left(\rho_{S}^{(+)}(t),\rho_{S}^{(-)}(t)\right)<0.
	\end{equation}
	Any revival of this quantity that leads to its change of sign (from negative to positive) can be related to a back flow of information from the environment \cite{Breuer2009, Breuer2010}. In the case at hand here, $D_{S}\left(\rho_{S}^{(+)}(t),\rho_{S}^{(-)}(t)\right)$ coincides with the difference between the coherence of the two states of the system since the initial states are orthogonal between them  ~\cite{Breuer2009, Breuer2010, Apollaro2011, Baumgratz_2014} [cf.  the Supplemental Material (SM) available at~\cite{SM} for a detailed derivation of this]. 
	Thus, in the remainder of this work, we will expoit the behavior of the degree of distinguishability between the chosen initial states of $S$ to relate the information exchanged between system and environment to the variation of the quantum resource embodied by coherence~\cite{Adesso2017}.
	
	Fig.~\ref{fig:Dist.Tr Sis_Amb e derivada} (a) shows the dynamics of the trace distance in Eq.~\eqref{TD_sisAmb}, while Fig.~\ref{fig:Dist.Tr Sis_Amb e derivada} (c) (black dashed line) shows the trend followed by $\sigma_S(t)$ (time has been rescaled as $t=J_E\tau$, where $\tau$ is the actual evolution time and $J_E=700$ rad s$^{-1}$ is the coupling rate between environment qubits~\cite{Mendonca2020}). The oscillations that are visible in $D_S\left(\rho_{S}^{(+)}(t),\rho_{S}^{(-)}(t)\right)$, and the sign-switches showcased by $\sigma_S(t)$ suggest   a significant back-flow of information from the environment to the system. As we mentioned above, since $\sigma_{S}(t)$ is related to a coherence measure, such quantum resource can be the quantity transmitted to and recovered from the environment across the dynamics. To analyse such conjecture, in what follows we take the standpoint of the environmental system and verify the dynamics of resources.
	
	\noindent
	{\it Information flow.--}
	In order to verify the dynamics of the information inside the environment, we use a measure proposed in Ref.~\cite{Laine2010BF, Megier2021} (cf. Ref.~\cite{SM} for further details)
	\begin{equation}
		D_{E}\left(\rho_{E}^{(+)}(t),\rho_{E}^{(-)}(t)\right)= \left\|\rho_{E}^{(+)}(t)-\rho_{E}^{(-)}(t)\right\|_{1},
		\label{TD_ambiente}
	\end{equation}
	and its derivative, 
	$\sigma_{E}(t)$, 
	where $\rho_{E}^{(\pm)}(t)$ stands for the state of the environment at the a given instant of time $t$, conditioned to the initial state of the system be $\rho_{S}^{(\pm)}(0)$, thus obtained as $\rho_{E}^{(\pm)}(t)={\rm Tr}_S\left[\rho^\pm_{SE}(t)\right]$. 
	The behavior of Eq.~\eqref{TD_ambiente} is presented in Fig.~\ref{fig:Dist.Tr Sis_Amb e derivada} (b). The trace distance between conditional environmental states, in particular, has been studied by considering the elements of $E$ both collectively and individually (i.e. by addressing 
	each element in each chain as 
		{$\rho_{E,k}^{(\pm)}(t)=\text{Tr}_{\overline{k}}\left[\rho_E^{(\pm)}(t)\right]$, where $\text{Tr}_{\overline{k}}$ denotes the partial trace over all qubits in the environment except the $k^\text{th}$).} 
	Our simulations, which were performed by considering two seven-element chains, show a growing degree of distinguishability from an initial situation where the states of $E$ are perfectly indistinguishable. The maximum of Eq.~\eqref{TD_ambiente} is achieved when the Eq.~\eqref{TD_sisAmb} is minimum [cf. points A and B in Fig.~\ref{fig:Dist.Tr Sis_Amb e derivada} (b)], while the minimum of Eq.~\eqref{TD_ambiente} is achieved simultaneously to the maximum of Eq.~\eqref{TD_sisAmb} [cf. point C]. This is strongly suggestive of a process where information is transferred from $S$ to $E$ and back. As the environmental elements are arranged symmetrically in the two chains, the dynamics of the qubits of both chains are equivalent and, therefore, the results shown refer to the qubits of a single chain only. We also notice that, as time evolves, quantum coherence is transferred from a layer of environmental qubit to the next, until it reaches the end of the chain, approximately halfway through the plateau between points A and B. Then, the information goes back to the first qubit again.
	
	\begin{figure}[b]
		\centering
		\includegraphics[width=1.0\columnwidth]{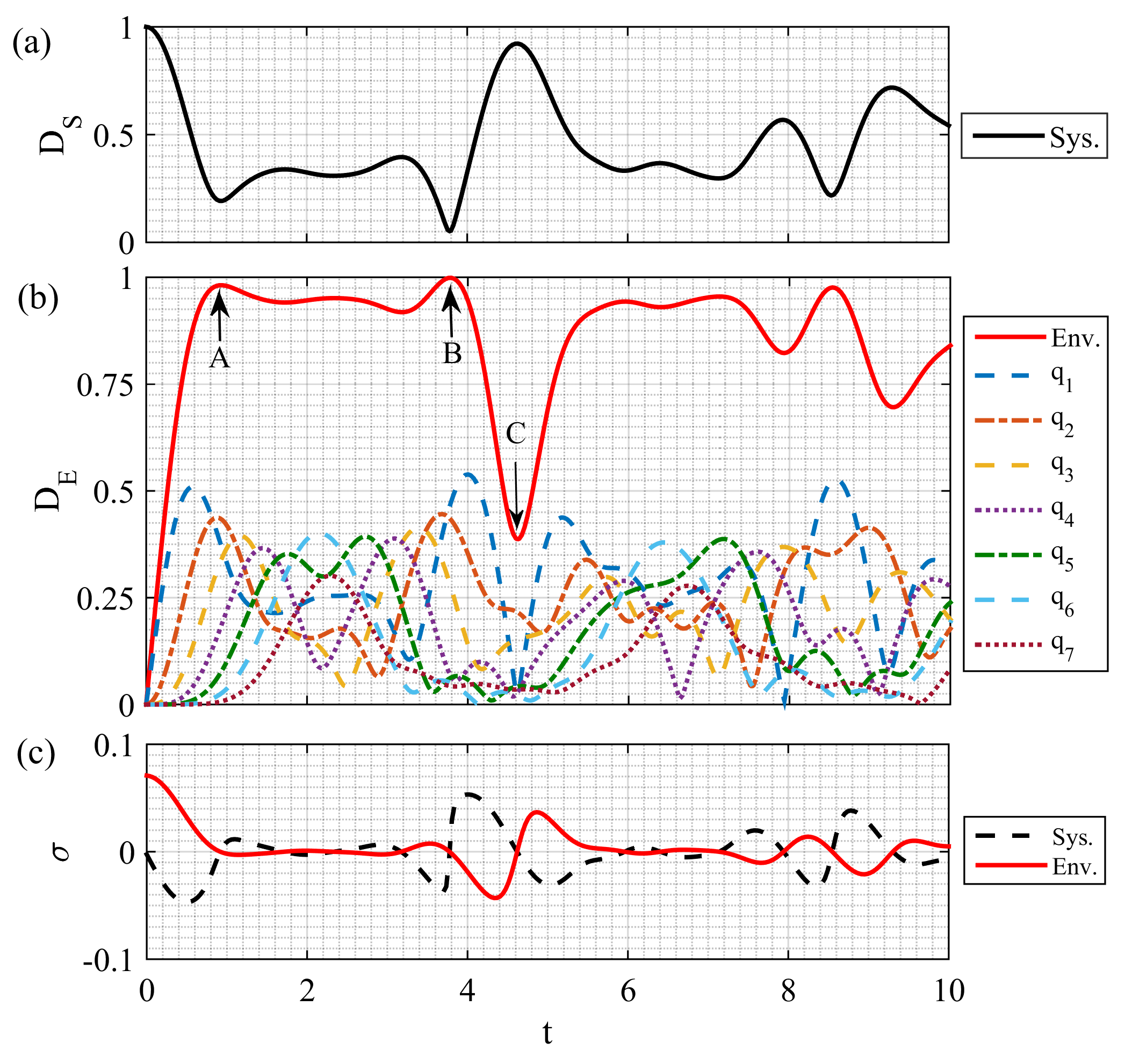}
		\caption{(a) Evolution of the trace distance between states $\rho_S^{(+)}(t)$ and $\rho_S^{(-)}(t)$. (b) Evolution of the trace distance between the states of the complete environment, $\rho_E^{(+)} (t)$ and $\rho_E^{(-)} (t)$, and between the states of each qubit of the chain. (c) Derivative in time of the trace distances between the states of the system and the states of the full environment. 
			Parameters used: $2N=14$ and $J_{SE}=0.71J_E$.}
		\label{fig:Dist.Tr Sis_Amb e derivada}
	\end{figure}
	
	The patterns identified above are quasi-periodic: when reaching point C, the process of information transfer is repeated: 
	the quantum coherence of the state of $S$ travels throughout the environments  and returns to the main system. The time needed to transfer most of the information from $S$ to $E$, i.e., to reach the plateau, is of the order of $t_{A}=1/J_{SE}$ and the time at which the plateau ends, i.e., the amount of time the information travels inside the environment (point B) is of the order of $t_{B}=N/2J_{E}$. Therefore, we have a relationship between sending information from the system qubit to the environment characterized in the system-environment coupling constant $J_{SE}$, whereas the coupling constant between the environment qubits $J_{E}$ establishes the information return. In Ref.~\cite{SM} we show the dynamics of our system against the coupling rates.
	
	As mentioned before, the violation of monotonic decrease in Eq.~\eqref{sigma} gives a signature of non-Markovian behavior of the dynamics \cite{Breuer2009, Breuer2010}. Fig.~\ref{fig:Dist.Tr Sis_Amb e derivada}(c) shows the evolution of $\sigma_S(t)$ (black dashed line) and $\sigma_E(t)$ (red solid line). It is interesting to note that both figures present plateaus of stability, suggesting that during the evolution, the information flow has ceased.

	\noindent
	{\it Link to Quantum Darwinism.--}
	The plateaux pinpointed in the analysis above suggest very strongly a potential connection with the selection of environment eigenstates and, thus, a link to quantum Darwinism~\cite{Zurek2009_Darwin,Zurek2013_Darwin}. Mutual information is a perfectly apt tool to quantify the information shared between $S$ and each of the {\it fragments} into which the environmental chains comprised in $E$ can be partitioned (as per Fig.~\ref{fig:Sistema}). 

\begin{figure}[t!]
	\centering
	\includegraphics[width=1.0\columnwidth]{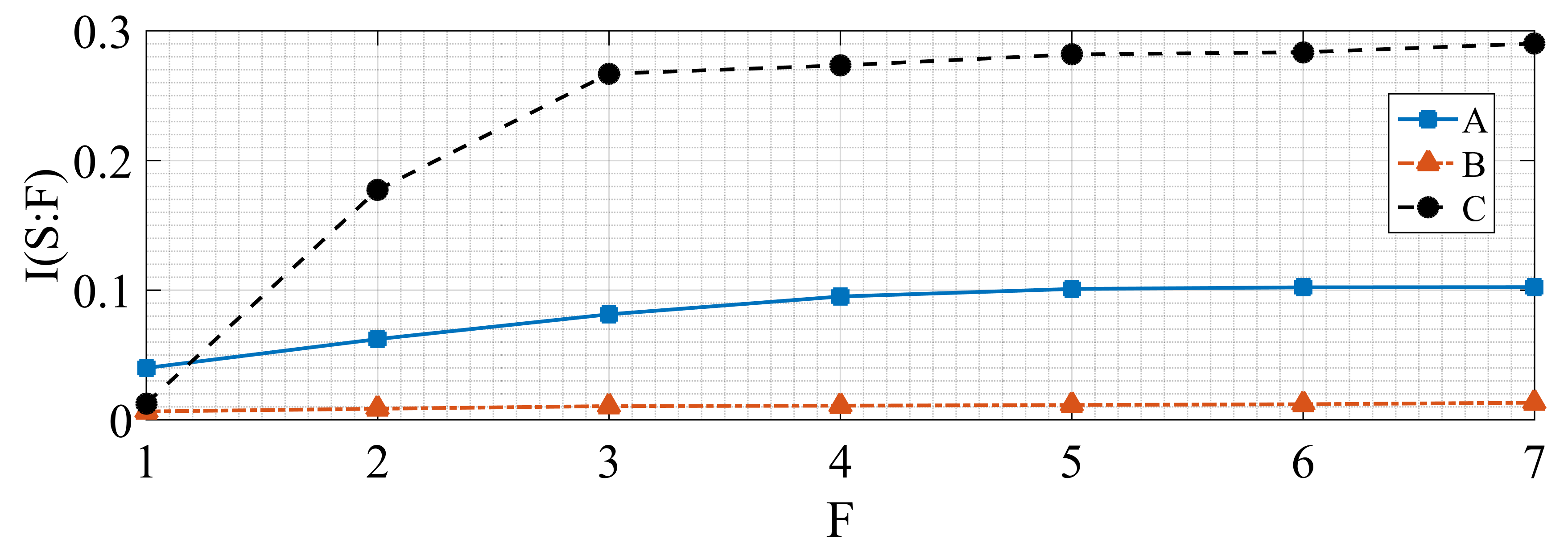}
	\caption{Mutual information at points A, B and C shown in Fig. \ref{fig:Dist.Tr Sis_Amb e derivada} for a whole system of 15 qubits.}
	\label{fig:IM}
\end{figure}

Unlike what we did previously, when we traced the environment out qubit by qubit, here we will trace different environment layers, thus building fragments of growing size. By doing this we will be able to retain the correlations between the main system and $E$, and explore the dynamics of information to different observers. Such correlations have been proven to be key in characterizing both the nature of the dynamics induced by finite-size environment such as the one addressed here and the phenomenology of quantum Darwinism~\cite{Vacchini2022a,Megier2021}.
Thus, we consider an environment divided into independent fragments of different sizes (Fig. \ref{fig:Sistema}),
exploring the possibility of achieving a plateau in the mutual information between the system and such fragment, which is a qualitative witness of the emergence of Darwinism~\cite{Ollivier_2001POVM, Zurek2006_Darwin, Mauro2019_Darwin1, Zurek2009_Darwin, Mauro2022_Darwin2, Darwin_Sebastian2022, Darwin_Sebastian2023}. 
While in Ref.~\cite{SM} we address the nature of the correlations under scrutiny, here we report on the behavior of the mutual information 
\begin{equation}
	I\left(\rho_{S}:\rho_{F_m}\right)= \mathcal{S}(\rho_S)+\mathcal{S}(\rho_{F_m})-\mathcal{S}(\rho_{SF_m}).
\end{equation}
where $\mathcal{S}(\rho)= -\text{tr}(\rho \text{log}\rho)$ is the von Neumann entropy 
~\cite{NielsenLivro2010} and $F_m$ stands for the $m^\text{th}$ fragment ($F_m$ thus comprises the elements with $k=1,\dots,m$ of both chains with $m=1,\dots,N$).


In  Fig.~\ref{fig:IM} we show $I\left(\rho_{S}:\rho_{F_m}\right)$ against $m$ for $N=7$ and at points A (information travels to the environment), B (backflow of information to the system), and C (instant of time at which the proces is complete and starts over) in Fig.~\ref{fig:Dist.Tr Sis_Amb e derivada}. 
While for A and B only a partial Darwinistic behavior is achieved, when C is considered and $m=N$, we have $I(S:E)=2 S(\rho_S)$~\cite{Zurek2013_Darwin}, signalling a full establishment of objective reality stemming from system-environment interaction, and reinforcing the qualitative link with the behavior of the information exchanged by $S$ and $E$ at the core of the discussion in the previous Section.

\noindent
{\it Conclusions.--}
We have studied the dynamics of information transfer between a system and its environment in a compound inspired by an adamantane molecule-based NMR setting~\cite{Mendonca2020,Alvarez2010,Ajoy2011,Souza2011}.
We have been able to provide a very detailed description of the phenomenology of such transfer, pinpointing the time at which information is sent to the environment and back. Remarkably, the occurrence of dynamical plateaux in the trace distance that quantifies the degree of distinguishability between different input states of the system and characterizes the non-Markovianity of the ensuing dynamics allowed us the establishment of a qualitative link with the pheonmenon of redundant encoding of information that is at the core of Quantum Darwinism.
Such link has been made robust through the study of mutual information, which clearly highlighted the uniform spreading of information on $S$ across growing fractions of $E$.

The detailed characterization of such information transfer processes in an open-system dynamics will be key to understand the redistribution of energy and quantum coherence across a quantum compound, which could in turn help characterizing dynamical processes relevant, for instance, in quantum thermodynamics.

\noindent
{\it Acknowledgments.--}
T.M.M. acknowledges funding via Grants No. 2021/01277-2 and No. 2022/09219-4, S\~ao Paulo Research Foundation (FAPESP). D.O.S.P. acknowledges the support by the Brazilian funding agencies CNPq (Grant No. 304891/2022-3), FAPESP (Grant No. 2017/03727-0), and the Brazilian National Institute of Science and Technology of Quantum Information (INCT/IQ). M.P. acknowledges the support by the European Union’s Horizon Europe EIC Pathfinder project QuCoM (Grant Agreement No.101046973), the Leverhulme Trust Grant UltraQuTe (Grant No. RGP-2018-266), the Royal Society Wolfson Fellowship (RSWF/R3/183013), the U.K. EPSRC (EP/T028424/1), the Department for the Economy Northern Ireland under the U.S.-Ireland R\&D Partnership Programme, the "Italian National Quantum Science and Technology Institute (NQSTI)" (PE0000023)-SPOKE 2 through project ASpEQCt, and the National Centre for HPC, Big Data and Quantum Computing (HPC) (CN00000013)-SPOKE 10 through project HyQELM. 



%


\clearpage
\pagebreak
\widetext
\begin{center}
\textbf{\Large Supplemental Material} \\
\vspace{0.125cm}
\textbf{\large System-environment quantum information flow} \\ 
\vspace{0.25cm}
{Taysa M. Mendon\c{c}a$^{1,2}$, Lucas C\'eleri$^{3}$, Mauro Paternostro$^{4,2}$ and Diogo O. Soares-Pinto$^{1}$}
\end{center}

\begin{center}
	$^{1}$Instituto de F\'{i}sica de S\~{a}o Carlos,
	Universidade de S\~{a}o Paulo, CP 369, 13560-970, S\~{a}o Carlos, SP, Brazil
	
	$^{2}$Centre for Theoretical Atomic, Molecular, and Optical Physics, School of Mathematics and Physics, Queen's University, Belfast BT7 1NN, United Kingdom
	
	$^{3}$QPequi Group, Institute of Physics, Federal University of Goi\'as, Goi\^ania, GO, 74.690-900, Brazil
	
	$^{4}$Universit\`a degli Studi di Palermo, Dipartimento di Fisica e Chimica - Emilio Segr\`e, via Archirafi 36, I-90123 Palermo, Italy

\par\end{center}
\setcounter{equation}{0}
\setcounter{table}{0}
\setcounter{section}{0}
\setcounter{theorem}{0}
\makeatletter
\renewcommand{\thesection}{S.\Roman{section}} 
\renewcommand{\thesubsection}{S.\Roman{section}.\arabic{subsection}}
\def\@gobbleappendixname#1\csname thesubsection\endcsname{\Alph{section}.\arabic{subsection}}
\renewcommand{\theequation}{S\arabic{equation}}
\renewcommand{\thefigure}{\arabic{figure}}
\renewcommand{\thetheorem}{S\arabic{theorem}}
\renewcommand{\thedefinition}{S\arabic{definition}}
\renewcommand{\bibnumfmt}[1]{[#1]}
\renewcommand{\citenumfont}[1]{#1}
\renewcommand{\bibnumfmt}[1]{[S#1]}
\renewcommand{\citenumfont}[1]{S#1}




\section{Trace distance and quantum coherence}
\label{sec:Mark_coer}

Considering two orthogonal states $\rho_{1}(t)$ and $\rho_{2}(t)$ whose dynamics consist of the same unital map

\begin{equation}
	\rho_{1}(t) = \frac{1}{2}(\mathbb{I}+\Vec{r}_1(t)\cdot\Vec{\sigma})= \frac{1}{2}
	\begin{pmatrix}
		1+z_1(t) & x_1(t)-\text{i}y_1(t)\\
		x_1(t)+\text{i}y_1(t) & 1-z_1(t)
	\end{pmatrix}
\end{equation}
and
\begin{equation}
	\rho_{2}(t) = \frac{1}{2}(\mathbb{I}+\Vec{r}_2(t)\cdot\Vec{\sigma})= \frac{1}{2}
	\begin{pmatrix}
		1+z_2(t) & x_2(t)-\text{i}y_2(t)\\
		x_2(t)+\text{i}y_2(t) & 1-z_2(t)
	\end{pmatrix}.
\end{equation}

We will use the definition of trace distance as
\begin{equation}
	\label{TD_sisAmb}
	D\left(\rho_{1}(t),\rho_{2}(t)\right)= \left\|\text{A}\right\|_{1},
\end{equation}
with $\text{A}=\rho_{1}(t)-\rho_{2}(t)$ and $ \| \text{A}\|_{1} = \frac{1}{2}\mbox{Tr}\left(\sqrt{\text{A}^{\dagger}\text{A}}\right)$, we have (the time dependence is omitted for the sake of simplicity)
\begin{equation}
	A=A^{\dagger}=\rho_{1}-\rho_{2}= \frac{1}{2}
	\begin{pmatrix}
		z_1-z_2 & (x_1-x_2)-\text{i}(y_1-y_2)\\
		(x_1-x_2)+\text{i}(y_1-y_2) & -z_1+z_2
	\end{pmatrix}.
	\end{equation}
	
	Thus $\sqrt{\text{A}^\dag \text{A}}=\vert \text{A}\vert$ so that $\|\text{A}\|_1=\sum^2_{i=1}\vert\lambda_i\vert$ with $\lambda_{1,2}$ the eigenvalues of $\text{A}$. For the two states given above, such eigenvalues take the form 
	\begin{equation}
		\lambda_1=-\lambda_2=\sqrt{(x_1-x_2)^2+(y_1-y_2)^2+(z_1-z_2)^2}=\abs{\Vec{r}_1-\Vec{r}_2},
	\end{equation}
	where we have used the condition $\text{Tr}[\rho_{1,2}]=1$ and nothing else. Thus, we can write the trace distance between two states through its vectors of Bloch sphere 
	\begin{equation}
		D(\rho_1,\rho_2)=\frac{1}{2}\abs{\Vec{r}_1-\Vec{r}_2}.
	\end{equation}
	
	If we use two orthogonal states such that $z_1(0)=z_2(0)=0$, as in our case, we can still find the trace distance is directly linked the difference between the coherence elements. 
	
	
	\section{Information Flow}
	\label{sec:Information Flow}
	
	To study the exchange of information between the system and the environment and to quantify the distinguishability between quantum states, we will analyze the inequality proposed by Laine \textit{et al.} in Ref.~\cite{Laine2010BF}, which has been the object of several studies~\cite{Megier2021, Vacchini_2019, Vacchini_2022}. Such inequality reads
	\begin{align}
		D\left(\rho_{S}^{(+)}(t),\rho_{S}^{(-)}(t)\right){-} D\left(\rho_{S}^{(+)}(\tau),\rho_{S}^{(-)}(\tau)\right)\leq D\left(\rho_{E}^{(+)}(\tau),\rho_{E}^{(-)}(\tau)\right){+}  
		\sum_{j=\pm} D\left(\rho_{SE}^{(j)}(\tau),\rho_{S}^{(j)}(\tau) \otimes \rho_{E}^{(j)}(\tau) \right)
		\label{DT_completo}
	\end{align}
	such that $t$ is greater than the one at time $\tau$. The initial conditions are $\rho_{SE}^{(\pm)}(0)=\rho_{S}^{(\pm)}(0) \otimes \rho_{E}^{(\pm)}(0)$, $\rho_{E}^{(+)}(0)=\rho_{E}^{(-)}(0)$. 
	
	The left-hand side of Eq.~\eqref{DT_completo}  quantifies the information, gained or lost, by the  system  in the time interval $t-\tau$. 
	The first term on the right side quantifies the distinguishability between the evolved states of the environment $\rho_{E}^{(\pm)}(t)$. As $\rho_{E}^{(+)}(0)=\rho_{E}^{(-)}(0)$, the dynamics of the term in question will depend on the information that leaves the main system going to environment.
	The other two terms of Eq. \eqref{DT_completo} quantify correlations between the system and the environment during the evolutions of $\rho_{SE}^{(\pm)}(t)$.
	
	Fig.~\ref{Fig:Eq_Dist.Traco} shows the evolution of Eq.~\eqref{DT_completo} for the full system described in Sec.~\textit{Model and resulting dynamics} of the main text.
	The dotted black line shows the dynamics of the right-hand side of Eq.~\eqref{DT_completo}, that is the dynamic of the system in a reduced amount of time. Note that there are regions very close to zero, these regions correspond to zero information flow, that is, all information leaving the system returns to it. The dot-dashed red line shows $D\left(\rho_{E}^{(+)}(t),\rho_{E}^{(-)}(t)\right)$, which is initially identically zero because the initial states of the environment  are completely indistinguishable.
	However, as the system evolves, the trace distance of environmental states increases until they become completely distinguishable. such a region is equivalent to the region where the leftmost term of Eq.~\eqref{DT_completo} is null.
	Therefore, the information coming from the system is being transferred to the environment and back again. This becomes clear when we look at the dynamics of individual environment qubits, as we did in Fig. 2(b) of the main text.
	
	The  solid blue and  and dashed green lines in Fig.~\ref{Fig:Eq_Dist.Traco} refer to the last term of Eq.~\eqref{DT_completo}. The correlations involving $\rho_{SE}^{(+)}(t)$ and $\rho_{SE}^{(-)}(t)$ are identical because of the symmetry of the system. Such lines reveal that there is an amount of information shared between the main system and the environment. 
	For identical initial states of the environment, an increase in the trace distance of the states of the environment indicates that there is an accumulation of correlations between $S$ and $E$ during the dynamics, such quantity is fundamental for the sharing of quantum resources between quantum objects, thus enabling the occurrence of the information exchange dynamics characteristic of the system + bath set.

	\begin{figure}[ht]
		\centering

		\includegraphics[width=1.0\columnwidth]{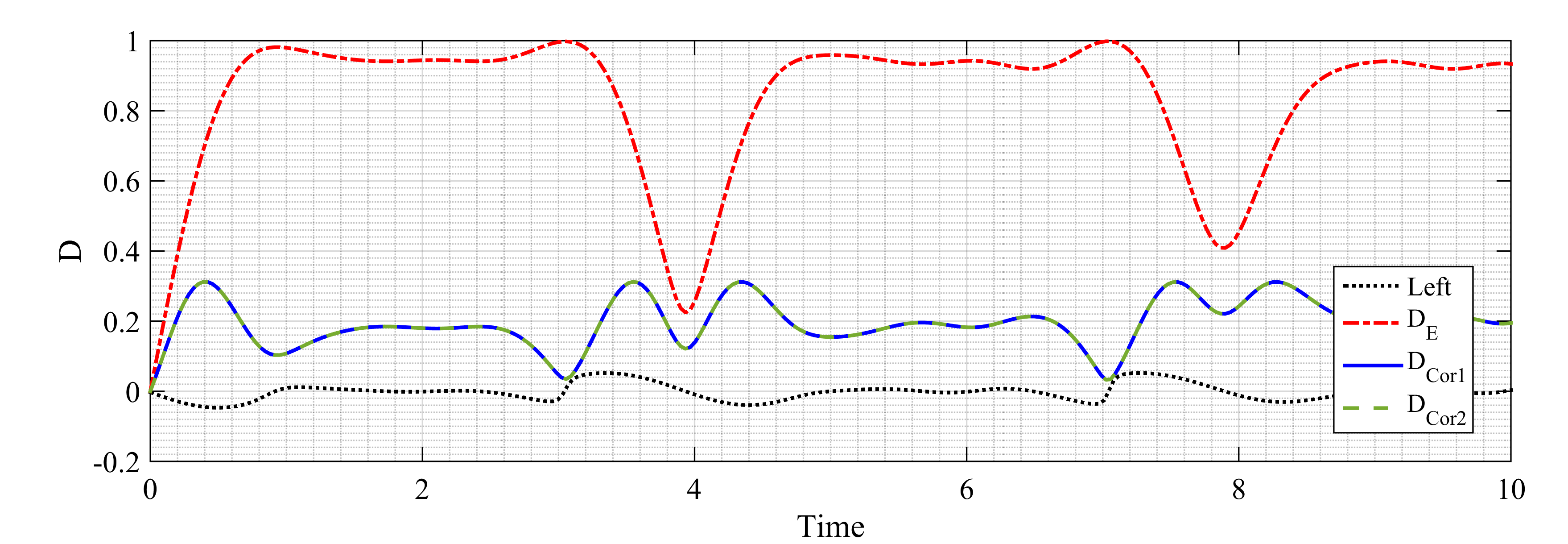}
		
		\caption{Illustration of the terms in Eq.~\eqref{DT_completo}. The dotted black line represents the left-hand side of Eq.~\eqref{DT_completo}, the dot-dashed red line refers to the first term on the right-hand side, while the solid blue and dashed green lines refer to its last two terms. We have used $2N=12$ qubits in environment and $J_{SE}=0.71J_E$. The horizontal axis shows the evolution time in dimensionless unit, as per the main text.}
		\label{Fig:Eq_Dist.Traco}
	\end{figure}
	
	\section{Markovianity measure and parameters of the full system}
	\label{sec:VariandoJ}
	
	Motivated by Ref.~\cite{Mendonca2020}, we now show how the transfer of information  from/to the main system to/from the environment is  influenced by parameters such as the coupling rates  and the number of elements of the environment.
	In Fig.~\ref{Fig:VariandoJ} we show how the trace distance between states of the environment (Eq.~(4) of the main text), and thus also that between states of the system, is influenced when we vary the number $N$ of elements qubits of each chain and the coupling rates  $J_{SE}$ and $J_E$. 
	
	In Fig. \ref{Fig:VariandoJ}(a) we kept the coupling between the main system and the qubits of the environment $J_{SE}$ fixed and we varied the coupling between the qubits of the environment $J_E$.
	We can see that at $T_{A}\simeq1/J_{SE}$, point highlighted by the arrow, the stability of the thermalization plateau begins, this point is fixed and does not depend on the number of qubits in the environment or $J_E$.
	The other end of the plateau is governed by the number of qubits in the environment and the coupling between them such that $T_{B}\simeq N/(2J_{E})$, as shown in Fig. \ref{Fig:VariandoJ}(b) where we fixed $J_E$ and vary $J_{SE}$.
	
	Therefore, we have a relationship between the measure of Markovianity, couplings - $J_{SE}$ and $J_E$ - and the number of qubits $N$ of each chain. However, this relationship has not been shown to work when $J_{SE} \simeq J_E$ or when $J_{SE} \ll J_E$.
	
	\begin{figure}[ht]
		\centering
		\includegraphics[width=0.9\columnwidth]{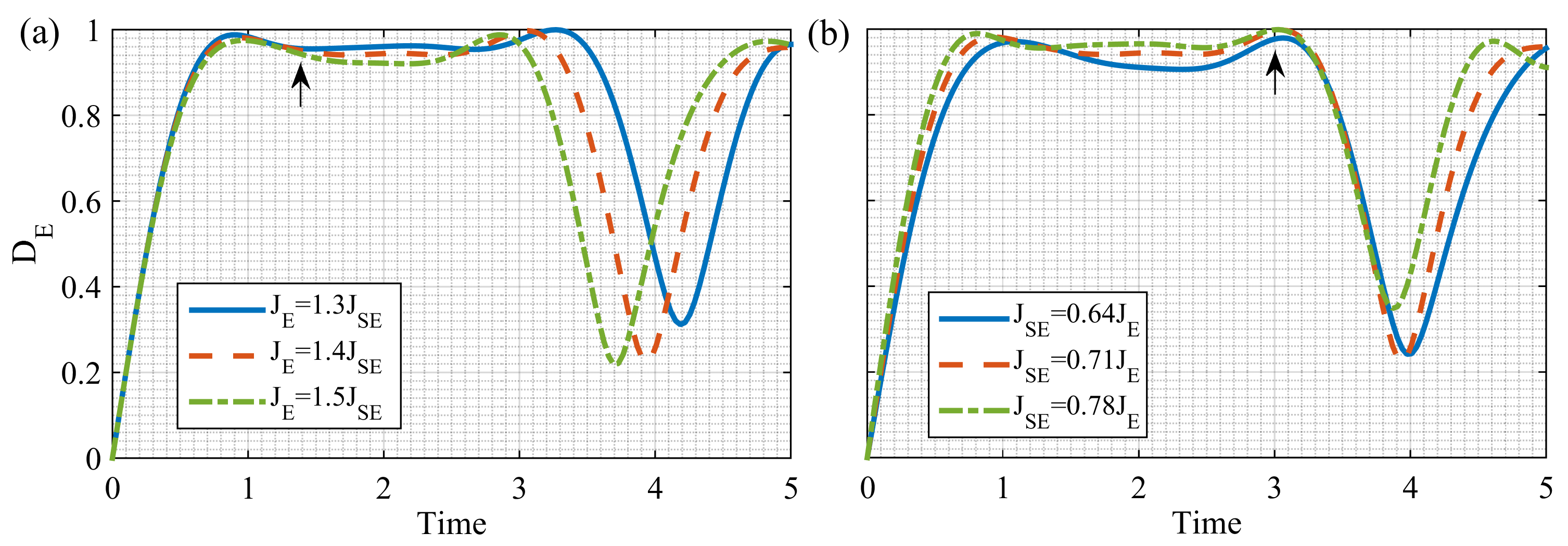}
		
		\caption{Trace distance between states of the environment (a) fixing $J_{SE}$ and varying $J_{E}$ and (b) fixing $J_{E}$ and varying $J_{SE}$ for a whole system containing 13 qubits.}
		\label{Fig:VariandoJ}
	\end{figure}
	
	\section{Strong Quantum Darwinism}
	
	In the main text we show that the system studied has regions where we can find the same information of the main system contained in different fragments of the environment, therefore Quantum Darwinism manifests itself in our model. This phenomenon appears stronger as we increase the number of qubits in the environment, as shown in Fig. \ref{Fig:IMxF} for (a) 10, (b) 12 and (c) 14 qubits in environment. The points blue, red and black are referents to regions A, B and C, respectively, showed on Fig. 2(b) of the main text.
	
	\begin{figure}[ht]
		\centering
		\includegraphics[width=0.8\columnwidth]{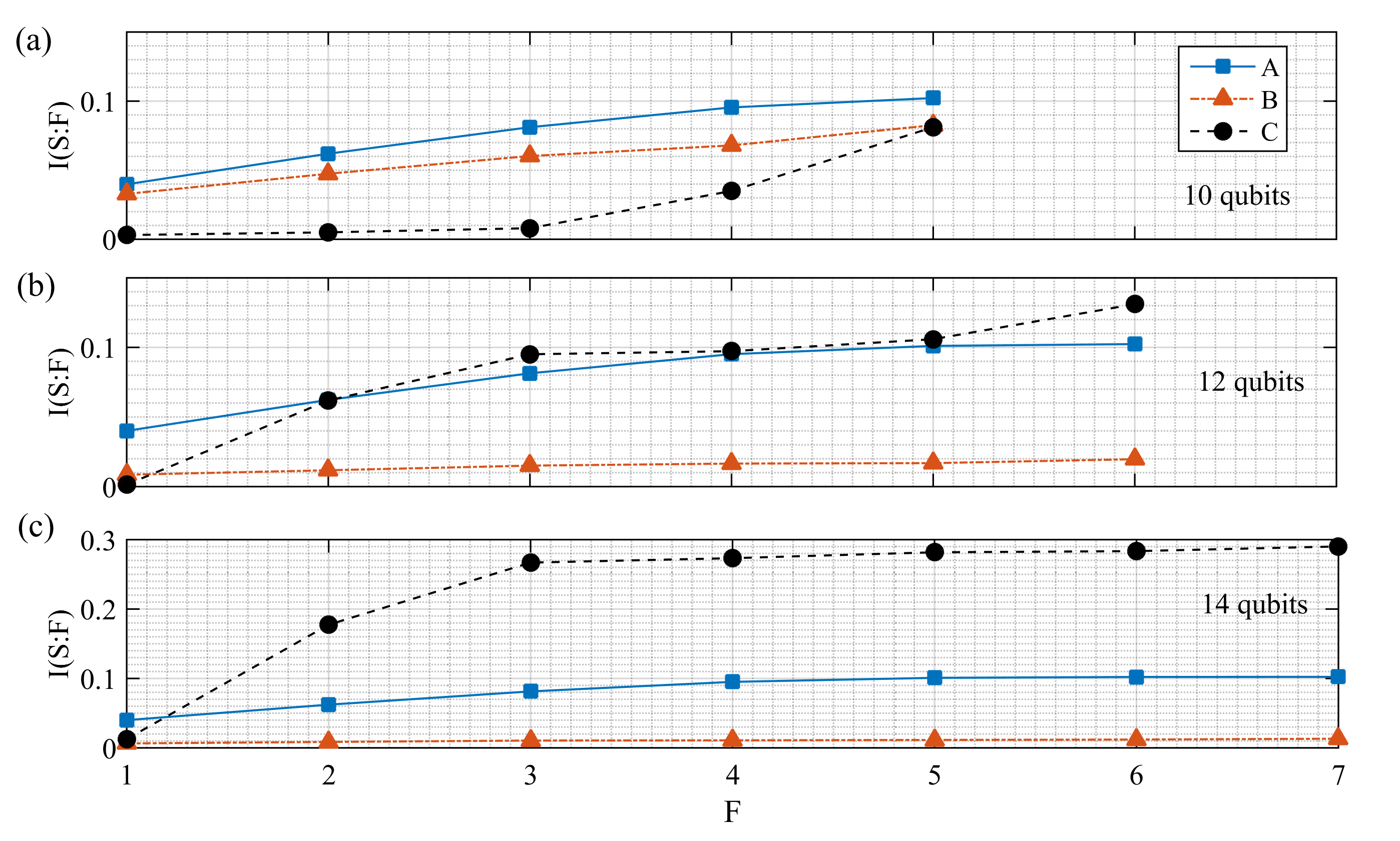}
		\caption{Mutual information \textit{versus} fragment size of environments with (a) 11, (b) 12 and (c) 14 qubits.}
		\label{Fig:IMxF}
	\end{figure}
	
	In addition to the points of interest mentioned, Fig. \ref{Fig:IMxt}(a) shows that we can find the same value of mutual information between the system and different fragments, that is, there is a manifestation of Quantum Darwinism in other regions. However, the information measured in these regions comes almost exclusively from the sending of information between the system and the first qubit of each chain, as shown in Fig. 2(c) of the main text, this fact is evident in Fig. \ref{Fig:IMxt}(b) where we show the mutual information between the system and each qubit of a chain.
	We can also see the dynamics of the information flow between the qubits of the environment in Fig. \ref{Fig:IMxt}(b), we see that the information of the system flows qubit by qubit between the qubits of the chain, this strengthens the result of the measurement of the evolution of the trace distance between the states of the states of each qubit of the chain shown in Fig. 2(b) of the main text.

	\begin{figure}[ht]
		\centering
		\includegraphics[width=0.8\columnwidth]{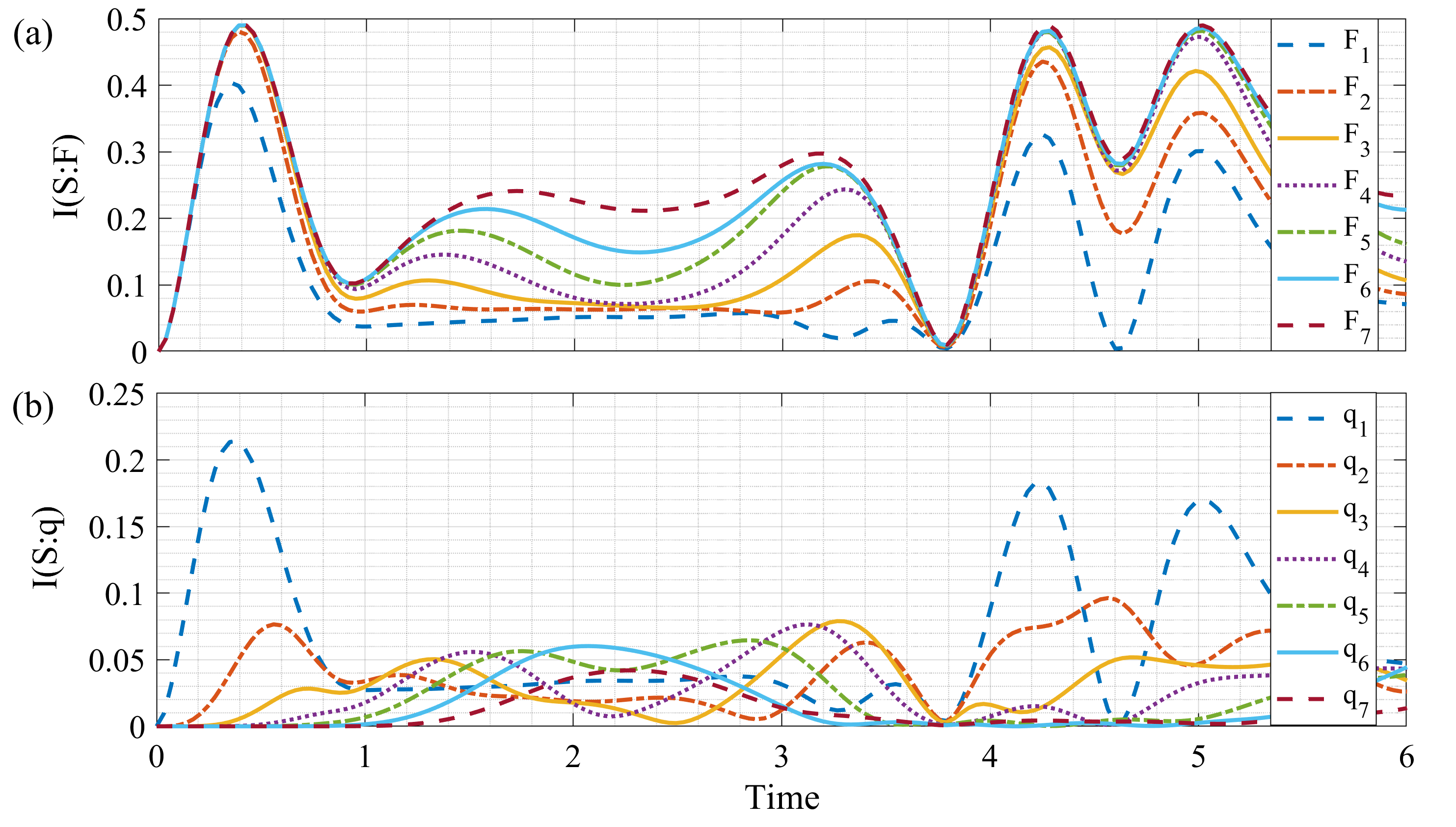}
		\caption{Temporal evolution of Mutual information. Mutual information between the system and (a) fragments of the environment and (b) qubits of one of the chains of the environment.}
		\label{Fig:IMxt}
	\end{figure}

	To complement the study of quantum Darwinism, in this section we will analyze whether strong quantum Darwinism \cite{Castro2019_StrongDarwin} occurs in our system.
	Strong quantum Darwinism assumes that there is no correlation between different fragments of the environment - see Fig. 1 of the main text - so the nature of the observed information is just classical.
	
	The quantum discord between main system and environment is defined as \cite{Zurek2009_Darwin,Ollivier_2001POVM}
	\begin{equation}       
		\mathcal{D}\left(\rho_{E}|\rho_S\right)=I\left(\rho_{S}:\rho_{E}\right) - \mathcal{J}(\rho_E|\rho_S)
	\end{equation}
	where $I\left(\rho_{S}:\rho_{E}\right)= \mathcal{S}(\rho_S)+\mathcal{S}(\rho_{E})-\mathcal{S}(\rho_{SE})$ is the mutual information (Eq. ($8$) of the main text) and 
	$\mathcal{J}(\rho_E|\rho_S)=\mathcal{S}(\rho_E)-{\text{min}_{\left\{\Pi_k\right\}}}\sum_{k}p_{k}\mathcal{S}\left(\rho_{E|k}\right)$ is the Holevo information. Here, the measure is performed on system $S$, we applied a Positive Operator Valued Measure (POVM) with elements $\Pi_k=M_k^{\dagger}M_k$, where $M_k=\left\{\cos\theta \ket{0}+e^{i\phi}\sin\theta \ket{1},e^{-i\phi}\sin\theta \ket{0}+\cos\theta \ket{0}\right\}$ is the measurement operator and a is the classical outcome \cite{Ollivier_2001POVM}.
	The probability of measuring the main system will be $p_k=\text{tr}(\Pi_k\rho_{SE})$ and the conditional state of the environment $\rho_{E|k}=\text{tr}_S(\Pi_k\rho_{SE})/p_k$ \cite{Modi_2012Discord}.
	
	Thus, the quantum discord for different fractions of the environment has the form:
	\begin{equation}
		\mathcal{D}\left(\rho_{F_m}|\rho_S\right)=\text{min}_{\left\{P_a\right\}}\sum_{a}p_{a}\mathcal{S}\left(\rho_{F_m|a}\right)+\mathcal{S}\left(\rho_S\right)+\mathcal{S}\left(\rho_{SF_m}\right).
	\end{equation}
	
	In analyzing quantum discord, we analyze the nature of information. When the quantum discord is different from zero, the correlations between the system and the environment have a quantum nature. If this is zero, only classical information is available on different fragments of the environment, this is the characteristic of strong quantum Darwinism.
	
	Fig. \ref{Fig:Darwin} shows the result of the simulation of quantum discord for a system with 14 qubits in the environment.
	In the case where we have full environment we have $\mathcal{D}\left(\rho_{E}|\rho_S\right)= \mathcal{S}\left(\rho_S\right)$, and consequently $I(S:E) = 2 D(S:F_m)$, as predicted by  Zwolak and Zurek \cite{Zurek2013_Darwin}.
	In the region referring to point B - point referring to the return of information from the environment to the main qubit - we see that the discord is very close to zero, but it does not signal strong Darwinism since it still has elements of quantum correlation. Therefore, we conclude that our system has Quantum Darwinism but not characteristic of Strong Quantum Darwinism.
	
	\begin{figure}[ht]
		\centering
		\includegraphics[width=1.0\columnwidth]{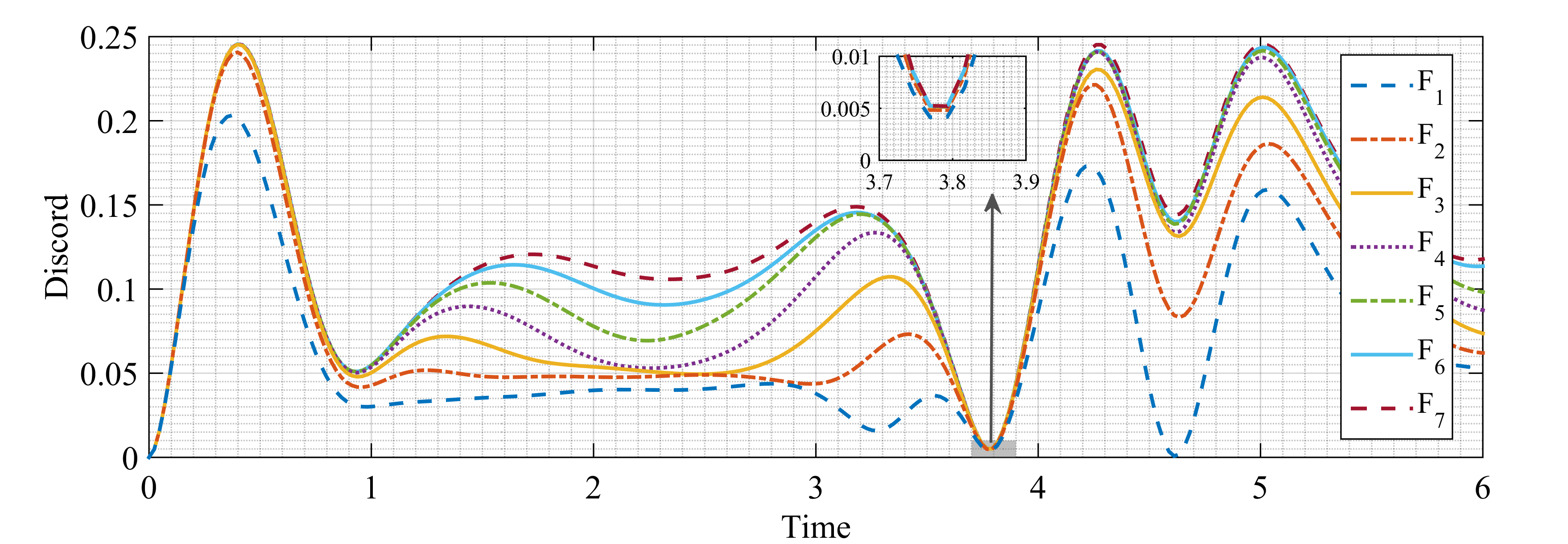}
		\caption{Quantum Discord for the fragments from a full system with 15 qubits.}
		\label{Fig:Darwin}
	\end{figure}



\end{document}